\documentclass[11pt]{article}
\usepackage{moriond,epsfig}

\def\be{\begin{equation}}
\def\ee{\end{equation}}
\def\bea{\begin{eqnarray}}
\def\eea{\end{eqnarray}}

\newcommand{\afig}{{\kern -0.25em a}}
\newcommand{\bfig}{{\kern -0.25em b}}
\newcommand{\cfig}{{\kern -0.25em c}}
\newcommand{\dfig}{{\kern -0.25em d}}

\newcommand{\MW}{M_{\mathrm{W}}}

\newcommand{\MH}{M_{\mathrm{H}}}

\newcommand{\TO}{\rightarrow}
\newcommand{\EE}{\mathrm{e^+e^-}}
\newcommand{\WW}{\mathrm{W^+W^-}}

\newcommand{\FFFF}{\mathrm{ffff}}

\newcommand{\QQP}{\mathrm{q} \overline{\mathrm{q^\prime}}}
\newcommand{\QQ}{\mathrm{q} \overline{\mathrm{q}}}
\newcommand{\LN}{\ell \nu}

\newcommand{\QQLN}{{\mathrm{q q} \ell \nu}}
\newcommand{\QQQQ}{\mathrm{q q q q}}

\def\GeV{\ifmmode {\mathrm{\ Ge\kern -0.1em V}}\else
                   \textrm{Ge\kern -0.1em V}\fi}%
\def\MeV{\ifmmode {\mathrm{\ Me\kern -0.1em V}}\else
                   \textrm{Me\kern -0.1em V}\fi}%
\def\keV{\ifmmode {\mathrm{\ ke\kern -0.1em V}}\else
                   \textrm{ke\kern -0.1em V}\fi}%
\def\eV{\ifmmode  {\mathrm{\ e\kern -0.1em V}}\else
                   \textrm{e\kern -0.1em V}\fi}%
\begin{document}
\vspace*{4cm}
\title{QCD AND W MASS DETERMINATION AT LEP}

\author{ ARNO STRAESSNER }

\address{EP Division, CERN, 1211 Geneva 23, Switzerland \\
e-mail: Arno.Straessner@cern.ch}

\maketitle\abstracts{ Measurements of hadronic final state interactions
  in W-pair events at LEP are presented. In particular, different
  scenarios of colour reconnection are studied. The data agree with
  moderate colour reconnection effects. Extreme scenarios are ruled
  out, limiting the systematic uncertainty on the LEP W mass
  determination due to this effect. An updated measurement of the mass
  of the W boson at LEP yields $80.412\pm 0.042 \GeV$.}

\section{Introduction}
At LEP, W bosons are produced in the reaction $\EE \TO \WW$ with the
subsequent decay of the W's into quark pairs, $\QQP$, or a lepton and
a neutrino, $\LN$. About 10000 W pair events are registered by each of
the four experiments ALEPH, DELPHI, L3 and OPAL, corresponding to a
total luminosity of $4\times700$ $\mathrm{pb}^{-1}$.

One of the main goals of the LEP programme is to determine the mass of
the W boson, $\MW$, from the reconstructed invariant mass spectra. 
In this measurement, the systematic uncertainty turns out to be
comparable to the statistical precision that can be reached. 
Especially in the fully hadronic decay channel, $\WW \TO \QQQQ$, the
presence of final state interactions (FSI), like colour
reconnection~\cite{cr} (CR) and Bose-Einstein correlations~\cite{bec}
(BEC), may cause a cross-talk and momentum transfer between the
reconstructed W bosons, which leads to a bias in the measured W mass.
Therefore, both effects, CR and BEC, are studied in more detail. The
latter one is discussed elsewhere~\cite{bec} in these
proceedings.

\begin{figure}
\hfill
\epsfig{figure=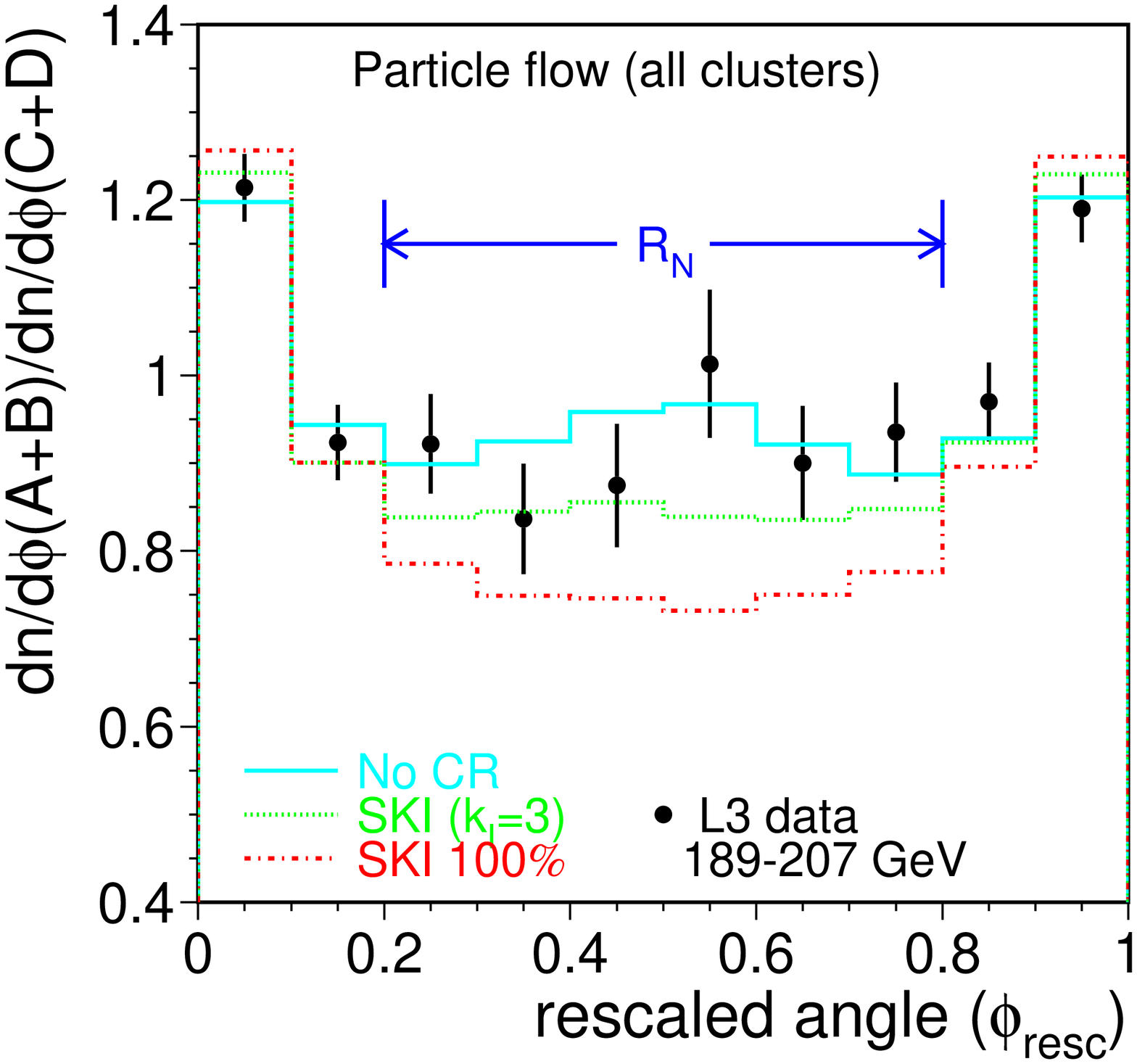,width=0.44\linewidth}
\hfill
\epsfig{figure=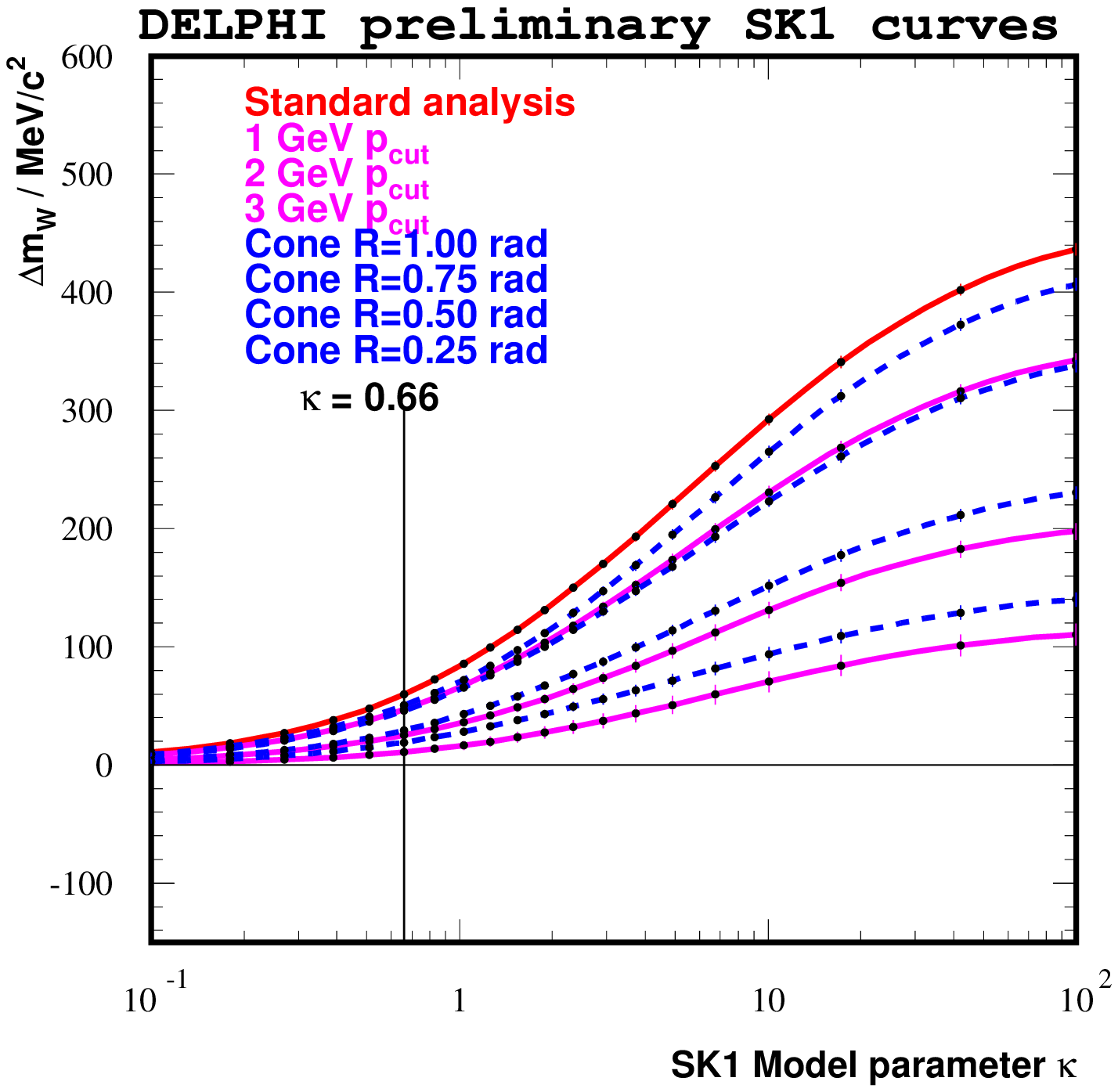,width=0.44\linewidth}
\hfill
\caption{a) Ratio of the particle flow $dn/d\phi$ in the inter-W
  (A+B) and intra-W (C+D) regions as measured by the L3 experiment and
  compared to MC predictions with and without CR.  The CR measure $R_N$
  is determined in the interval indicated by the arrow. b) Applying
  stronger cuts on particle momenta or on jet cone sizes reduces the W
  mass bias predicted by the SK1 model.
  \label{fig:cr}}
\end{figure}

\section{Colour Reconnection}
In $\WW \TO \QQQQ$ events, the two quarks originating from each W
boson form a colour singlet. In absence of FSI, the fragmentation
processes of the quark pairs, measured as hadronic jets in the
detector, are independent. Energy-momentum is separately conserved in
the two W systems. However, in case of colour reconnection the
momentum distribution of hadrons is rearranged. In the string picture
of fragmentation this is due to a reconnection of strings between the
original colour singlets.  Several Monte Carlo (MC) models~\cite{cr}
exist to describe the CR effect in W pair events: the
Sj{\"o}strand-Khoze (SKI, SKII) and generalised-area-law (GAL) models
implemented in Pythia, and the CR models of Ariadne (AR2) and Herwig
(HW). The GAL model is not investigated here since there are strong
limits on the size of its CR effect from Z peak data~\cite{rathsman}.

In the default configuration, all models predict a fraction of
reconnected events of about 30\%. In the SKI model, which is commonly
used as a benchmark, this fraction can continuously be varied. The
event reconnection probability is calculated according to
$P_{reco}=1-\exp(-k_I f)$, where $k_I$ is a free parameter and $f$ the
overlap integral of the strings that are identified with colour flux
tubes in the SKI model.

A measurement~\cite{aleph-multi} of the charged particle multiplicity,
$N_{ch}$, in $\QQQQ$ events is found not to be sensitive enough to CR.
ALEPH measures a multiplicity difference in $\QQQQ$ and $\QQLN$ events
of $\Delta_{ch}=N_{ch}(\QQQQ)-2N_{ch}(\QQLN)=0.31\pm 0.23 (stat.)\pm
0.10 (syst.)$, to be compared with a maximal effect from CR of
$\Delta_{ch}=0.46\pm0.03$ in the Herwig model. All other models predict even
smaller values of $\Delta_{ch}$. The analysis of the particle momentum
distribution yields that differences between the fragmentation models
are larger than the expected CR effect, even in the soft momentum
regime.

The string reconnection picture proposes that sensitivity is gained by
studying the ratio of the particle flow in the region between the jets
of each hadronically decaying W (intra-W) and in the region between
jets of different W's (inter-W).  To determine the flow in the two
regions, the particle momenta are projected on planes that are spread
between the directions of the four jets, where the first two jets
belong to one W and the last two jets to the second W. The angles,
$\phi$, of the particles to the jet directions are normalised such
that the rescaled angular difference between two neighbouring jets is
one.  Figure~\ref{fig:cr}~\afig\ shows the measured ratio of the
particle flow $dn/d\phi$ in the intra-W and inter-W regions. The SKI
model predicts a clear depletion of the flow in the central angular
interval.

To quantify the CR effect, the ratio $R_N=\int_{0.2}^{0.8}
dn/d\phi(\mathrm{intra-W})d\phi / \int_{0.2}^{0.8}
dn/d\phi(\mathrm{inter-W})d\phi $ is calculated and normalised to the
expectation without CR, $r=R_N^{data}/R_N^{no-CR}$. Combining the
results of the LEP experiments and taking their different CR
sensitivities due to different W pair purities into account, the
following values of $r$ are found for LEP data~\cite{cr-lep} and MC:
\begin{eqnarray*}
  r_\mathrm{AR}^\mathrm{LEP} = 0.959 \pm 0.010 (stat.)\pm 0.010 (syst.)\,,& & r^\mathrm{MC}_\mathrm{AR2\, CR} = 0.989 \,;\\
  r_\mathrm{HW}^\mathrm{LEP} = 0.950 \pm 0.011 (stat.)\pm 0.010 (syst.)\,,& & r^\mathrm{MC}_\mathrm{HW\, CR} = 0.987 \,;\\
  r_\mathrm{SKI}^\mathrm{LEP} = 0.969 \pm 0.011 (stat.)\pm 0.011 (syst.)\,,& & r^\mathrm{MC}_\mathrm{SKI 100\%\, CR} = 0.891 \,.
\end{eqnarray*}
The main systematic uncertainties are due to hadronisation modelling
($\pm 0.008$) and $\EE \TO \QQ\mathrm{gg}$ background shape and scale ($\pm
0.003$).  Data deviates from the AR2 and HW CR predictions by $-2.1
\sigma$ and $-2.6 \sigma$, respectively.  The extreme 100\% SKI
scenario can be excluded at $5.2 \sigma$. However, the agreement with
the no-CR scenarios is in all cases only at a level of about $-2
\sigma$. When the SKI parameter $k_I$ is left free, the best agreement
with data is found for $k_I=1.18$, corresponding to a reconnection
probability of 49\%. The 68\% confidence level lower and upper
limits are 0.39 and 2.13, respectively. Thus, data favours a moderate
reconnection fraction, as it is modelled in SKI. However, in the case
of AR2 and HW no conclusion can be drawn, because these models do not
predict significant changes in the colour flow.

\begin{figure}
\hfill
\parbox{0.45\linewidth}{
\footnotesize
\begin{tabular}{l | r | r | r }
    Source & \multicolumn{3}{c}{Uncertainties}
      \\
      & \multicolumn{3}{c}{ on $\MW$ in $\MeV$} \\
    & $\QQLN$ & $\QQQQ$ & \ $\FFFF$\ \\
    \hline
     Colour Reconnection &   -- &    90 &        9 \\
     BE Correlations     &   -- &    35 &        3 \\
     Hadronisation       &   19 &    18 &       18 \\
     ISR/FSR               &      8 &       8 &          8 \\
     Detector Syst.        &     14 &      10 &         14 \\
     LEP Beam Energy       &     17 &      17 &         17 \\
     Other                 &      4 &       5 &          4 \\
    \hline
    Total Systematic             &   31 &   101 &       31 \\
    \hline
    Statistical                  &   32 &    35 &       29 \\
    \hline
     Total                   &    44 &     107 & 
      43  \\
    \hline
    Statistical in ab- &   32 &    28 &       21 \\
    sence of Systematics  & & & \\
\end{tabular}
\normalsize
}
\hfill
\parbox{0.44\linewidth}{
  \epsfig{figure=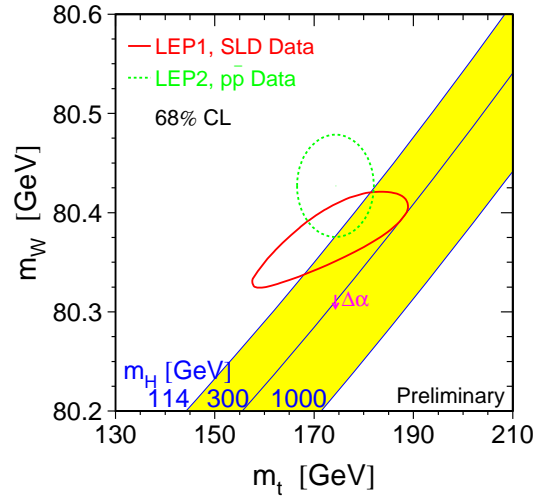,width=1.0\linewidth}
}
\hfill
\caption{a) Systematic uncertainties in the W mass measurement. b)
  Contour curves of the direct and indirect determinations of the W
  boson and Top quark masses, together with the predictions of the
  Standard Model for different values of the Higgs boson mass, $\MH$. Direct and indirect
  measurements prefer small values of $\MH$.
  \label{fig:wmass}}
\end{figure}

\section{Measurement of the W Boson Mass}

The W boson mass is determined in a maximum likelihood fit by
comparing data to MC samples with different underlying W masses. The W
mass variation is implemented by means of MC reweighting or
convolution techniques. The most sensitive W mass estimator is the
average invariant mass, which is calculated from the reconstructed
final state fermions. A kinematic fit, imposing energy-momentum
conservation and equal invariant masses of the two decaying W's,
improves the resolution on the measured quantities.

With the current techniques a comparable statistical precision~\cite{mw-lep} on
$\MW$ is reached in the $\QQLN$ and $\QQQQ$ channel, of 32 and 35
$\MeV$, respectively. However, using the measured limits on the SKI
parameter $k_I$, the systematic uncertainty due to CR is ranging
between 75 and $105~\MeV$ in the hadronic channel, increasing with
centre-of-mass energy. This reduces the weight of this channel in the
combined W mass result to just 11\%.

Since recently, alternative W mass estimators are investigated that
are based on modified jet reconstruction algorithms. By removing low
momentum particles or restricting the cone size of the quark jets the
CR bias is found be reduced in the SKI model, which is illustrated in
Figure~\ref{fig:cr}~\bfig. This observation is turned into a
measurement of CR. By comparing the W mass bias in data and MC as a
function of momentum or cone cut, the CR scenario that fits best to
data can be determined. The DELPHI CR measurement~\cite{cr-delphi}
improves from $k_I=2.4^{+14.6}_{-2.2}$ using only the colour flow
method to $k_I=1.96^{+2.60}_{-1.30}$ combining it with the result
using the alternative mass estimator. The results of both methods are
practically uncorrelated because different event samples are used in
the analyses. Also the correlation with the W mass measurement is
found to be only about 10\%. The analyses by the other LEP experiments
are still in preparation.  Therefore, the current W mass measurements
do not yet include information from the alternative W mass estimators.

The current statistical and systematic uncertainties on the W mass are
given in Figure~\ref{fig:wmass}~\afig. In the hadronic channel,
uncertainties on CR and BEC dominate. For the combined results the
fragmentation uncertainties become important because they are
correlated between the $\QQLN$ and $\QQQQ$ channels.  These
uncertainties are derived from comparisons between different
fragmentation models~\cite{frag-models}, Pythia, Ariadne and Herwig,
and from a variation of the model parameters, that are tuned with Z
peak data. The uncertainty due to the LEP beam energy is expected to
be reduced with the final energy calibration.

Finally, the results on the LEP combined W mass measurements~\cite{mw-lep} are
\begin{eqnarray*}
\MW(\QQLN) &=& 80.411~\pm~0.032~(stat.)~\pm~0.030~(syst.)~\GeV\\
\MW(\QQQQ) &=& 80.420~\pm~0.035~(stat.)~\pm~0.101~(syst.)~\GeV\\
\MW(\FFFF) &=& 80.412~\pm~0.029~(stat.)~\pm~0.031~(syst.)~\GeV\,,
\end{eqnarray*}
where the $\QQLN$ and $\QQQQ$ results are 18\% correlated. The mass
difference between the $\QQQQ$ and $\QQLN$ channels is found to be $
+22\pm 43\MeV$, neglecting FSI uncertainties. This indicates that
there are no extreme CR effects present in data. However, a better
understanding of the QCD effects in the hadronic W decay is needed to
improve the current W mass result. Combining the LEP result with the
$\MW$ measurements at $p\bar{p}$ colliders, yields~\cite{mw-all}
$\MW=80.426\pm0.034\GeV$.

The mass of the W boson is an important parameter in the Standard
Model of electroweak interactions. In Figure~\ref{fig:wmass} the
direct measurements of the masses of the W boson and the Top quark~\cite{mtop} are
compared to the values that are determined in a Standard Model
analysis~\cite{mw-all} of all remaining LEP and SLD data, probing the
Standard Model at the level of its radiative corrections. Good
agreement between direct and indirect measurements is found. Also
shown is the Standard Model prediction for different values of the
Higgs boson mass. Both data sets prefer small values of $\MH$, close
to the current lower limit from Higgs boson searches~\cite{higgs-lep} at LEP.

\section*{Acknowledgements}
I would like to thank the experiments ALEPH, DELPHI, L3 and OPAL for
making their most recent and preliminary results available. I also
like to thank the LEP Electroweak and WW Working Groups for preparing
the results in form of nice graphs and plots.

\end{document}